\documentclass[9pt,conference,a4paper]{IEEEtran}

\usepackage{amssymb} 
\usepackage{amsfonts}
\usepackage{amsmath} 
\usepackage{graphicx} 

\def\Psf{{\mathsf{P}}}

\def\xbf{{\mathbf{x}}}
\def\ybf{{\mathbf{y}}}

\def\ybf{{\mathbf{y}}}

\def\sbf{{\mathbf{s}}}

\def\Ccal{{\mathcal{C}}}
\def\Dcal{{\mathcal{D}}}
\def\Rcal{{\mathcal{R}}}

\def\xbfhat{{\widehat{\mathbf{x}}}}

\def\R{{\mathbb{R}}}

\def\defn{{\,\triangleq\,}}
\def\argmin{\mathop{\mathsf{arg\,min}}}
\def\fix{\mathsf{fix}}
\def\prox{\mathsf{prox}}
\def\denoise{\mathsf{denoise}}

\begin{document}

\title{Plug-In Stochastic Gradient Method}

\author{%
\IEEEauthorblockN{
Yu Sun\IEEEauthorrefmark{1},
Brendt Wohlberg\IEEEauthorrefmark{2},
and Ulugbek~S.~Kamilov\IEEEauthorrefmark{1}\IEEEauthorrefmark{3}}
\IEEEauthorblockA{\IEEEauthorrefmark{1}
Department of Computer Science and Engineering, Washington University in St.~Louis, MO 63130, USA}
\IEEEauthorblockA{\IEEEauthorrefmark{2}
Los Alamos National Laboratory, Theoretical Division, Los Alamos, NM 87545, USA}
\IEEEauthorblockA{\IEEEauthorrefmark{3}
Department of Electrical and Systems Engineering, Washington University in St.~Louis, MO 63130, USA}
}

\maketitle

\begin{abstract}
Plug-and-play priors (PnP) is a popular framework for regularized signal reconstruction by using advanced denoisers within an iterative algorithm. In this paper, we discuss our recent online variant of PnP~\cite{Sun.etal2018a} that uses only a subset of measurements at every iteration, which makes it scalable to very large datasets. We additionally present novel convergence results for both batch and online PnP algorithms.
\end{abstract}

\section{Introduction}

Consider the problem of estimating an unknown signal $\xbf \in \R^n$ from a set of noisy measurements measurements $\ybf \in \R^m$. This task is often formulated as an optimization problem
\begin{equation}
\label{Eq:Optimization}
\xbfhat = \argmin_{\xbf \in \R^n} \left\{\Ccal(\xbf)\right\} \quad\text{with}\quad \Ccal(\xbf) = \Dcal(\xbf)+\Rcal(\xbf)\; ,
\end{equation}
where $\Dcal$ is the data-fidelity term and $\Rcal$ is the regularizer promoting solutions with desirable properties. Some popular regularizers in imaging include nonnegativity, sparsity, and self-similarity.

Two common algorithms for solving optimization problem~\eqref{Eq:Optimization} are fast iterative shrinkage/thresholding algorithm (FISTA)~\cite{Beck.Teboulle2009a} and alternating direction method of multipliers (ADMM)~\cite{Afonso.etal2010}. These algorithms are suitable for solving large-scale imaging problems due to their low-computational complexity and ability to handle the non-smoothness of $\Rcal$. Both ISTA and ADMM have modular structures in the sense that the prior on the image is only imposed via the proximal operator defined as
\begin{equation}
\prox_{\gamma \Rcal}(\ybf) = \argmin_{\xbf \in \R^n}\left\{\frac{1}{2}\|\xbf-\ybf\|_{\ell_2}^2 + \gamma \Rcal(\xbf)\right\} \;.
\end{equation}
The mathematical equivalence of the proximal operator to regularized image denoising has recently inspired Venkatakrishnan et al.~\cite{Venkatakrishnan.etal2013} to replace it with a more general denoising operator $\denoise_\sigma(\cdot)$ of controllable strength $\sigma > 0$. The original formulation of this \emph{plug-and-play priors (PnP) framework} relies on ADMM, but it has recently been shown that it can be equally effective when used with other proximal algorithms. For example,
\begin{subequations}
\label{Eq:PnPISTA}
\begin{align}
\label{Eq:PnPISTA1}&\xbf^k \leftarrow \denoise_\sigma(\sbf^{k-1}- \gamma \nabla \Dcal(\sbf^{k-1}))\\
\label{Eq:PnPISTA2}&\sbf^k \leftarrow \xbf^k + ((q_{k-1}-1)/q_k)(\xbf^k - \xbf^{k-1}) \;,
\end{align}
\end{subequations}
where $q_k \leftarrow 1$ and ${q_k \leftarrow \frac{1}{2}(1+\sqrt{1+4q_{k-1}^2})}$ lead to PnP-ISTA and PnP-FISTA, respectively, and $\gamma > 0$ is the step-size~\cite{Kamilov.etal2017} .

In many applications, the data-fidelity term consists of a large number of component functions
\begin{equation}
\Dcal(\xbf) = \frac{1}{k}\sum_{i = 1}^k \Dcal_i(\xbf) \quad\Rightarrow\quad \nabla \Dcal(\xbf) = \frac{1}{k}\sum_{i = 1}^k \nabla \Dcal_i(\xbf) \;,
\end{equation}
where each $\Dcal_i$ and $\nabla \Dcal_i$ depend only on a subset of the measurements $\ybf$. Hence, when $k \geq 1$ is large, traditional batch PnP algorithms may become impractical in terms of speed or memory requirements. The central idea of the recent PnP-SGD method~\cite{Sun.etal2018a} is to approximate the full gradient in~\eqref{Eq:PnPISTA1} with an average of $b \ll k$ component gradients
\begin{equation}
\hat{\nabla} \Dcal(\xbf) = \frac{1}{b} \sum_{j = 1}^b \nabla \Dcal_{i_j}(\xbf)
\;,
\end{equation}
where $i_1, \dots, i_b$ are independent random variables that are distributed uniformly over $\{1, \dots, i\}$. The \emph{minibatch} size parameter $b \geq 1$ controls the number of per-iteration gradient components.

\begin{figure}[t]
\begin{center}
\includegraphics[width=8.5cm]{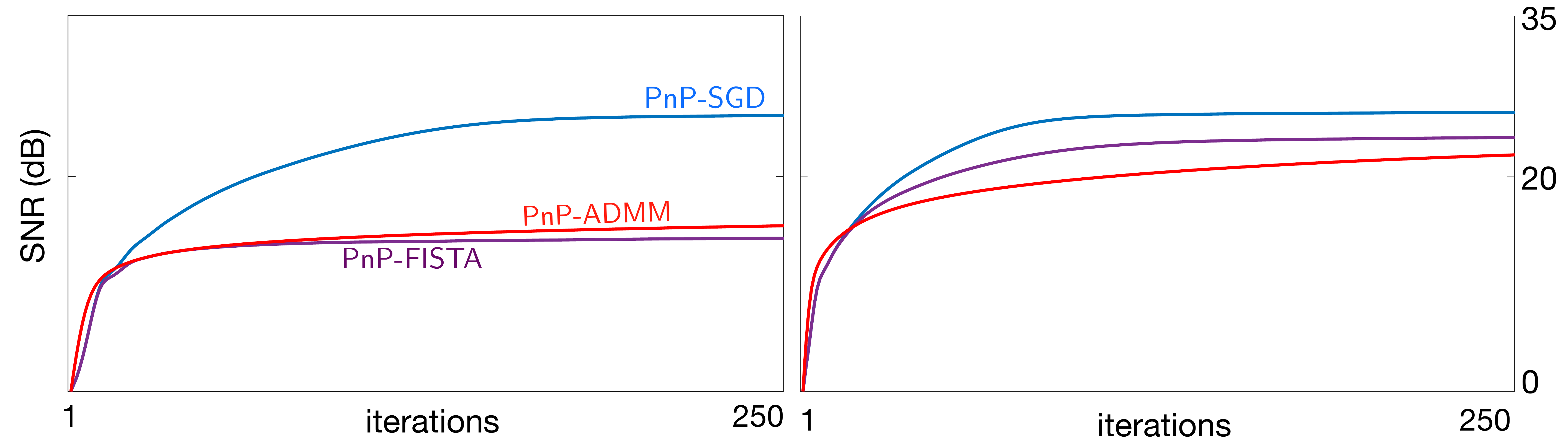}
\end{center}
\caption{Comparison between the batch and online PnP algorithms under a fixed measurement budget of 10 (left) and 30 (right). SNR (dB) is plotted against the number of iterations for three algorithms: PnP-SGD, PnP-FISTA, and PnP-ADMM. For the same per iteration cost, PnP-SGD can significantly outperform its batch counterparts. For a detailed discussion see~\cite{Sun.etal2018a}.}
\label{Fig:TestImages}
\end{figure}

We denote the denoiser-gradient operator by
\begin{equation}
\Psf(\xbf) \defn \denoise_\sigma(\xbf-\gamma \nabla \Dcal(\xbf))
\end{equation}
and its set of fixed points by
\begin{equation}
\fix(\Psf) \defn \{\xbf \in \R^n : \xbf = \Psf(\xbf)\} \;.
\end{equation}
The convergence of both batch PnP-FISTA and online PnP-SGD was extensively discussed in~\cite{Sun.etal2018a}. In particular, when $\Dcal$ is convex and smooth, and $\denoise_\sigma(\cdot)$ is an averaged operator (as well as other mild assumptions), it is possible to theoretically establish that the iterates of PnP-ISTA (without acceleration) gets arbitrarily close to $\fix(\Psf)$ with rate $O(1/t)$. Additionally, it is possible to establish that, in expectation, the iterates of PnP-SGD can also be made arbitrarily close to $\fix(\Psf)$ by increasing the minibatch parameter $b$. As illustrated in Fig.~\ref{Fig:TestImages}, this makes PnP-SGD useful and mathematically sound alternative to the traditional batch PnP algorithms.

\bibliographystyle{IEEEtran}

\begin{thebibliography}{1}
\providecommand{\url}[1]{#1}
\csname url@samestyle\endcsname
\providecommand{\newblock}{\relax}
\providecommand{\bibinfo}[2]{#2}
\providecommand{\BIBentrySTDinterwordspacing}{\spaceskip=0pt\relax}
\providecommand{\BIBentryALTinterwordstretchfactor}{4}
\providecommand{\BIBentryALTinterwordspacing}{\spaceskip=\fontdimen2\font plus
\BIBentryALTinterwordstretchfactor\fontdimen3\font minus
  \fontdimen4\font\relax}
\providecommand{\BIBforeignlanguage}[2]{{%
\expandafter\ifx\csname l@#1\endcsname\relax
\typeout{** WARNING: IEEEtran.bst: No hyphenation pattern has been}%
\typeout{** loaded for the language `#1'. Using the pattern for}%
\typeout{** the default language instead.}%
\else
\language=\csname l@#1\endcsname
\fi
#2}}
\providecommand{\BIBdecl}{\relax}
\BIBdecl

\bibitem{Sun.etal2018a}
Y.~Sun, B.~Wohlberg, and U.~S. Kamilov, ``An online plug-and-play algorithm for
  regularized image reconstruction,'' {\scriptsize\textsf{arXiv:1809.04693 [cs.CV]}}, 2018.

\bibitem{Beck.Teboulle2009a}
A.~Beck and M.~Teboulle, ``Fast gradient-based algorithm for constrained total
  variation image denoising and deblurring problems,'' \emph{IEEE Trans. Image
  Process.}, 2009.

\bibitem{Afonso.etal2010}
M.~V. Afonso, J.~M.Bioucas-Dias, and M.~A.~T. Figueiredo, ``Fast image recovery
  using variable splitting and constrained optimization,'' \emph{IEEE Trans.
  Image Process.}, 2010.

\bibitem{Venkatakrishnan.etal2013}
S.~V. Venkatakrishnan, C.~A. Bouman, and B.~Wohlberg, ``Plug-and-play priors
  for model based reconstruction,'' in \emph{Proc. {GlobalSIP}}, 2013.

\bibitem{Kamilov.etal2017}
U.~S. Kamilov, H.~Mansour, and B.~Wohlberg, ``A plug-and-play priors approach
  for solving nonlinear imaging inverse problems,'' \emph{IEEE Signal. Proc.
  Let.}, 2017.

\end{thebibliography}


\end{document}